\documentclass[aps,prl,twocolumn,superscriptaddress]{revtex4-2}

\usepackage{mathtools}
\usepackage[scaled=1.005]{newtxtext}
\usepackage{amsmath}
\usepackage{amsfonts}
\usepackage[scaled=1.005,varbb,varg,uprightGreek]{newtxmath}
\usepackage{bm,dsfont,eucal}
\AtBeginDocument{\renewcommand{\vec}[1]{\bm{#1}}}

\usepackage{graphicx}
\usepackage[dvipsnames,svgnames,x11names,hyperref]{xcolor}
\usepackage{hyperref}
\hypersetup{ 
    colorlinks=true, linkcolor = NavyBlue,
    urlcolor  = NavyBlue, citecolor = NavyBlue
}

\usepackage{physics,braket,siunitx,slashed}
\usepackage[capitalise]{cleveref}

\newcommand{\pdagger}{{\phantom{\dagger}}}
\newcommand{\cd}{c^\dagger}
\newcommand{\cpd}{c^\pdagger}
\newcommand{\ia}{i\alpha}

\newcommand{\smashop}{\smashoperator}
\newcommand{\dmu}{\delta\mu}
\newcommand{\dt}{\delta t}
\newcommand{\bs}[1]{\boldsymbol{#1}}
\newcommand{\K}{\boldsymbol{k}}

\begin{document}

\title{Non-unitary multiorbital superconductivity from competing interactions in Dirac materials}

\author{Tobias M. R. Wolf}
\affiliation{Institute for Theoretical Physics, ETH Zurich, 8093 Zurich, Switzerland}

\author{Maximilian F. Holst}
\affiliation{Institute for Theoretical Physics, ETH Zurich, 8093 Zurich, Switzerland}

\author{Manfred Sigrist}
\affiliation{Institute for Theoretical Physics, ETH Zurich, 8093 Zurich, Switzerland}

\author{Jose L. Lado}
\affiliation{Department of Applied Physics, Aalto University, 00076 Aalto, Espoo, Finland}

\date{\today}

\begin{abstract}
Unconventional superconductors represent one of the most intriguing quantum states of matter. 
In particular, multiorbital systems have the potential to host exotic
\emph{non-unitary} superconducting states. 
While the microscopic origin of non-unitarity is not yet fully solved, competing interactions are suggested to play a crucial role in stabilizing such states. 
The interplay between charge order and superconductivity has been a recurring theme in unconventionally superconducting systems, ranging from cuprate-based superconductors to dichalcogenide systems and even to twisted van der Waals materials. 
Here, we demonstrate that the existence of competing interactions gives rise to a non-unitary superconducting state. 
We show that the non-unitarity stems from a competing charge-ordered state whose interplay with superconductivity promotes a non-trivial multiorbital order.
We establish this mechanism both from a Ginzburg-Landau perspective, and also
from a fully microscopic selfconsistent solution of a multiorbital Dirac material.
Our results put forward competing interactions as a powerful mechanism for driving non-unitary multiorbital superconductivity.
\end{abstract}

\maketitle

Materials with competing interactions represent a paradigmatic playground to engineer novel electronic states of matter \cite{Si2010,Andrei2021}. 
Generically, the interplay of different interaction channels, typically attractive electron-phonon coupling and repulsive Coulomb interactions, can give rise to competing states.
Such competitions lead to especially rich physics in the presence of many active electronic orbitals, where the existence of additional degrees of freedom substantially enlarges the space of emergent symmetry broken states \cite{PhysRevLett.94.147005,PhysRevB.94.104501}. 
Several materials show charge order coexisting with unconventional superconductivity, such as cuprate superconductors  \cite{Tranquada1995,PhysRevLett.75.4650,Sebastian2015}, two-dimensional materials \cite{Ugeda2015,Manzeli2017}, and twisted van der Waals heterostructures  \cite{Cao2018,Jiang2019}.
The latter have recently attracted much attention for their diverse and often tunable interplay between symmetry broken states \cite{Zondiner2020,Stepanov2020}.

Unconventional superconducting states  \cite{RevModPhys.63.239} are intensely pursued for their potential topological and exotic properties \cite{Sato2017}. 
Among them, non-unitary (NU) superconducting states take a special place \cite{PhysRevLett.71.625,Machida1998,PhysRevLett.118.016802,PhysRevLett.109.097001}.
In the multiorbital systems, NU superconductivity is characterized by non-trivial interorbital pairing that leads to a NU pairing matrix.
Generic mechanisms that ensure the emergence of these NU states are so far not well understood, yet tuning competing interactions between different orbitals has been shown to be a promising route
to stabilize complex interorbital pairing channels \cite{PhysRevB.70.054507,PhysRevB.88.045115}.
Finding and understanding minimal multiorbital models realizing this mechanism still remains an open problem in correlated quantum matter \cite{PhysRevB.78.064514,Fischer2013,PhysRevB.81.014511,PhysRevB.94.174513,PhysRevB.94.104501,PhysRevResearch.1.033107,PhysRevB.100.094504}.

\begin{figure}
\centering
\includegraphics[width=\columnwidth]{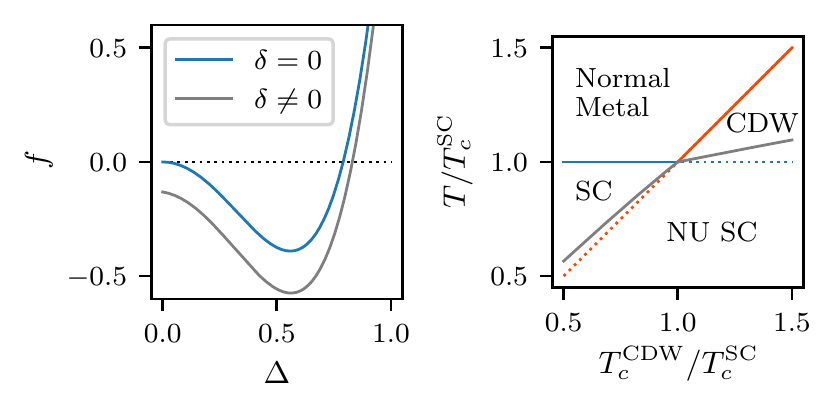}
\caption{%
Ginzburg-Landau theory for coupled order parameters $\Delta$ of unitary superconductivity (SC), $\delta$ of non-unitary SC (NU SC) and $m$ of a charge density wave (CDW) in a bipartite system.
(a)~Free energy density $f$ as a function of the SC order parameter $\Delta$, illustrating that the presence of a charge order ($\delta, m\neq 0$) lowers the condensation energy.
(b)~Phase diagram featuring regions where (non-unitary) SC and a CDW coexist. Shown are phase transition lines between a normal metal and a SC (CDW) state in absence of a CDW (SC) state in red (blue), as well as the transition line (gray) into the non-unitary phase where SC and the CDW coexist. We also show the case when the order parameters are uncoupled (dotted lines) [cf.~$\lambda=0$ in \cref{eq:GL_coupling}].
}
\label{fig:GL}
\end{figure}

In this Letter, we demonstrate that different interactions can cooperate to generate NU multiorbital superconductivity.
In particular, we first use a Ginzburg-Landau (GL) theory to show the cooperative effect between charge order and superconductivity (cf.~\cref{fig:GL}).
Secondly, we demonstrate -- using a microscopic model including both repulsive and attractive interactions -- how two distinct orders in the ground state of a Dirac system favor the onset of superconductivity with NU pairing. 
Interestingly, the emergence of the NU superconducting state from the cooperative charge and superconducting orders can be probed by the emergence of a gap opening away from the chemical potential, providing a simple experimentally observable signature of NU superconductivity.

We first address how the cooperative effect between a charge-density wave (CDW) and superconductivity (SC) can be captured using a symmetry analysis of the GL free energy of a multiorbital system. 
While the following argument does not depend on the existence of a Dirac crossing in the electronic spectra, this situation will be of particular interest when addressing a microscopic model realizing this phenomenology.
In the following, we consider a material with a bipartite lattice structure with two sublattices $A$ and $B$.
The corresponding point group $G$ contains symmetry operations $G'\subset G$ that preserve the two sublattices, as well as symmetry operations $G\setminus G'$ that exchange them.
In the bulk of this material, the SC state can be described by the GL free energy density
\begin{align}
    f_{\text{SC}} = a(T)\abs{\Delta}^2 + b\abs{\Delta}^4 ,
    \label{eq:GL_SC}
\end{align}
where $\Delta$ is the complex SC order parameter transforming according to an arbitrary one-dimensional irreducible representation ($1$D irrep) of the corresponding point group $G$. 
Close to the phase transition, we assume $a(T) = a'(T -T_c^{\text{SC}})$ and $b > 0$ for stability, where $T_c^{\text{SC}}$ is the bare SC critical temperature. 

Next, we consider the formation of a CDW that creates an imbalance in the bipartite sublattice structure. 
We express the electronic sublattice densities $n_A$ and $n_B$ through their average $n_0=(n_A+n_B)/2$ and difference $m=n_A-n_B$ to describe the CDW with the GL free energy density
\begin{align}
    f_{\text{CDW}} = \alpha(T) m^2 +\beta m^4 ,
    \label{eq:GL_CDW}
\end{align}
where $m$ is the real-valued CDW order parameter. The latter transforms according to the $1$D irrep of $G$ which has characters $+1$ ($-1$) on the conjugacy classes which preserve (exchange) the sublattices.
Similar to \cref{eq:GL_SC}, we choose $\alpha(T) = \alpha'(T -T_c^{\text{CDW}})$ and $\beta > 0$, where $T_c^{\text{CDW}}$ is the bare CDW critical temperature.

The imbalance in the electron densities of the two sublattices $A$ and $B$ caused by the CDW $m$ requires us to modify the free energy density in \cref{eq:GL_SC} to \footnote{
The higher-order terms for the stability of the free energy density are discussed in Ref.~\cite{supmat}.
}
\begin{align}
    f_{\text{SC}} = a_1(T)\left(\abs{\Delta_A}^2 +\abs{\Delta_B}^2\right) +a_2\abs{\Delta_A -\Delta_B}^2 ,
    \label{eq:GL_SC2}
\end{align}
where $\Delta_{A, B} = \abs{\Delta_{A, B}}e^{i\varphi_{A, B}}$ are the complex SC order parameters for each sublattice. 
The second term in \cref{eq:GL_SC2} is a coupling term that minimizes the phase difference between $\Delta_A$ and $\Delta_B$, allowing us to choose $\varphi_A=\varphi_B=0$.
Furthermore, we need to take into account the direct coupling between the SC and the CDW order parameters, i.e.,
\begin{align}
    f_{\text{CPL}} = \lambda m\left(\abs{\Delta_A}^2 -\abs{\Delta_B}^2\right) ,
    \label{eq:GL_coupling}
\end{align}
such that the total GL free energy density is given by
\begin{align}
    f = f_{\text{SC}} +f_{\text{CDW}} +f_{\text{CPL}} .
    \label{eq:GL_full}
\end{align}
We note that the coupling $f_{\text{CPL}}$ can naturally lead to $\abs{\Delta_{A}}\neq\abs{\Delta_{B}}$, which is a minimal sufficient condition to have a non-unitary pairing matrix.
Hence, we introduce the unitary SC order parameter $\Delta=(\Delta_A+\Delta_B)/2$, 
and the non-unitary (NU) superconducting order parameter $\delta=(\Delta_A-\Delta_B)/2$, to state \cref{eq:GL_full} in leading order as
\begin{align}
    f = 2a_1\Delta^2 + \begin{pmatrix} \delta & m \end{pmatrix}\begin{pmatrix} \frac{a_1}{2} +a_2 & \lambda\Delta \\ \lambda\Delta & \alpha \end{pmatrix}\begin{pmatrix} \delta \\ m \end{pmatrix}.
    \label{eq:GL_full2}
\end{align}
We study the stable phases and in particular the phase transition from the unitary SC phase (${\Delta\neq 0}$, ${m=\delta=0}$) to a NU phase with both SC and CDW order (${\Delta,m,\delta\neq 0}$), using the linearized GL equations 
\footnote{They are given by $\partial_{\delta}f =\partial_{m} f = 0$.},
i.e., we find the two eigenvalues of the matrix in \cref{eq:GL_full2}.

Generally, as shown in \cref{fig:GL}, we find that the coupling $f_{\text{CPL}}$ between non-unitary SC ($\delta$) and the charge-density wave ($m$) tends to be cooperative, i.e., SC and CDW favor each other.
In \cref{fig:GL}(a), we compare the free energy density for the superconducting phase in absence of the CDW (${\Delta\neq 0}$, ${\delta=m=0}$) and the non-unitary SC phase ($\Delta, \delta, m\neq0$).
We find that in the latter case the local minimum is lower and shifted towards larger magnitudes, indicating that the coupling to the CDW amplifies and stabilizes the SC order parameter $\Delta$.
In \cref{fig:GL}(b), we show a phase diagram for temperature $T$ versus critical temperature $T_c^{\text{CDW}}$ of the CDW.
For $T_c^{\text{SC}} > T_c^{\text{CDW}}$, the system is in the normal metal state above $T_c^{\text{SC}}$.
By lowering the temperature below $T_c^{\text{SC}}$, the system transitions into a unitary SC phase ($\Delta\neq0$ but $\delta=m=0$). 
A further reduction below $T_c$ leads to a second phase transition into NU SC order ($\Delta, \delta, m\neq0$).
Similarly, for $T_c^{\text{CDW}} > T_c^{\text{SC}}$, the system transitions from the normal metal phase first to a pure CDW phase and then to the NU SC phase.
We observe that the coupling between NU SC $\delta$ and the CDW $m$ increases the transition temperature from $T_c^{\text{CDW}}$ to $T_c>T_c^{\text{CDW}}$, which again signals that the coupling between order parameters mutually stabilises SC and CDW phases.

\begin{figure}
\includegraphics[width=\columnwidth]{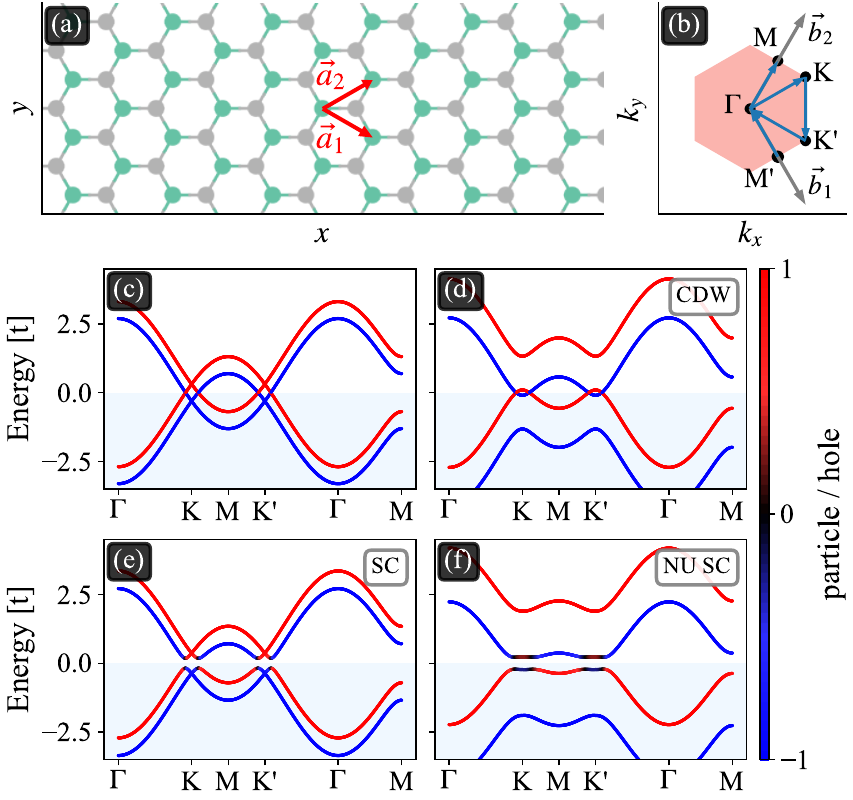}
\caption{ %
Honeycomb lattice and electronic band structures with and without mean-field interactions. 
(a) Real-space structure with lattice vectors $\vec{a}_1$, $\vec{a}_2$ and sublattices $\alpha=A,B$ (green, gray) and (b) reciprocal-space first Brillouin zone with high-symmetry points $\Gamma$, $M$, $K$, and $K'$ and high-symmetry path. 
(c-f) Energy bands along this high-symmetry path (c) in absence of interactions, (d) in a regime dominated by nearest-neighbor repulsion $V$ (CDW), (e) in a regime dominated by on-site attraction $U$ (SC), and (f) intermediate regime with both sizable interactions $U$ and $V$ (NU SC) [cf. stars highlighted in \cref{fig3}]. 
The band color distinguishes particle/hole branches (red/blue) that are required to treat particle-particle pairing.
}%
\label{fig2}
\end{figure}

After having observed the cooperative mechanism between CDW and SC from a phenomenological point of view, we move on to demonstrate that this behavior is also realized in a microscopic model.
As a minimal example featuring the mechanism of coupled order parameters, we consider a honeycomb lattice as shown in \cref{fig2}(a). 
Each unit cell contains two atoms with sublattice labels $\alpha\in\{A,B\}$ which we assume to be of same type. 
The point group is then $G=D_{6h}$ and when sublattice symmetry is broken the symmetry elements reduce to $G'=D_{3h}$.
In reciprocal space, the first Brillouin zone is also hexagonal and has high-symmetry points $\Gamma$, $K$, $K'$, and $M$, see \cref{fig2}(b). 

We can describe the electronic properties in the tight-binding approximation with the Hamiltonian
\begin{align} \label{eq:tightbinding_noninteracting}
  H_0 = \smashop{\sum_{\langle i A, j B\rangle,s}} t \, c_{i A s}^{\dagger} c_{j B s}^{\pdagger}+ \mathrm{h.c.}  - \mu \sum_{i,\alpha,s} \, n_{\ia s} \;,
\end{align}
where $t$ is the hopping amplitude between nearest neighbor orbitals, $\cpd_{\ia s}$ ($\cd_{\ia s}$) destroys (creates) an electron on sublattice $\alpha$ with spin ${s\in\{\uparrow,\downarrow\}}$ in unit cell $i$. 
The second term accounts for the chemical potential $\mu$ ($\mu=0$ at half-filling $\nu=0.5 $), where ${n_{\ia s}=c_{\ia s}^{\dagger} c_{\ia s}^{\pdagger}}$ is the local electron density operator. 
In \cref{fig2}(c), we see the electronic band structure of Hamiltonian \eqref{eq:tightbinding_noninteracting} along a high-symmetry path. 
It shows a characteristic Dirac crossing near the $K$ and $K'$ points. 
We assume that the chemical potential is slightly below half-filling ($\nu=0.5$) such that the lower electronic band is partially empty (here $\nu=0.49$) with circular Fermi surfaces around the $K$- and $K'$-point. 

We now include additional electronic interactions in the model. We consider short-range interactions of strength $U<0$ (attractive) between electrons on the same site and strength $V>0$ (repulsive) between nearest neighbors, i.e.,
\begin{align} \label{eq:extended_hubbard}
H_{\textrm{int}} = U \sum_{\ia} n_{\ia\uparrow} n_{\ia\downarrow} 
  + V \smashop{\sum_{\langle iA, jB\rangle, ss'}} n_{iA s} n_{jB s'}.
\end{align}
Using the mean field approximation, we find that the Hamiltonian in \cref{eq:tightbinding_noninteracting} gets modified as 
$t \to \overline{t}_{s s'}(G)$ 
and 
$\mu \to \overline{\mu}_{\alpha s}(G)$ 
through mean fields 
$G_{i\alpha,j\beta}^{ss'}=\langle \cd_{i\alpha s}\cpd_{j\beta s'}\rangle$
and 
$F_{i\alpha,j\beta}^{ss'}=\langle \cd_{i\alpha s}\cd_{j\beta s'}\rangle$,
where $\langle\cdot\rangle$ is the ground state expectation value of the corresponding mean-field Hamiltonian $\overline{H}_0$, see Ref.~\cite{supmat} for details.
The approximation also introduces particle-particle pairing terms, of which we include the on-site spin singlet contribution
\begin{align}\label{eq:mf_h_pairing}
  \overline{H}_{P} &= \frac{U}{2} \sum_{i,\alpha} \Delta_{i\alpha,i\alpha}^{0\;\ast} \, c^\dagger_{\ia s} (-i \sigma^y)_{ss'}^{\pdagger} c^\dagger_{\ia s'}
  + \textrm{h.c.},
\end{align}
where $\Delta_{i\alpha,i\alpha}^{0}(F)$ is the local singlet pairing amplitude.

\begin{figure}
\includegraphics[width=\columnwidth]{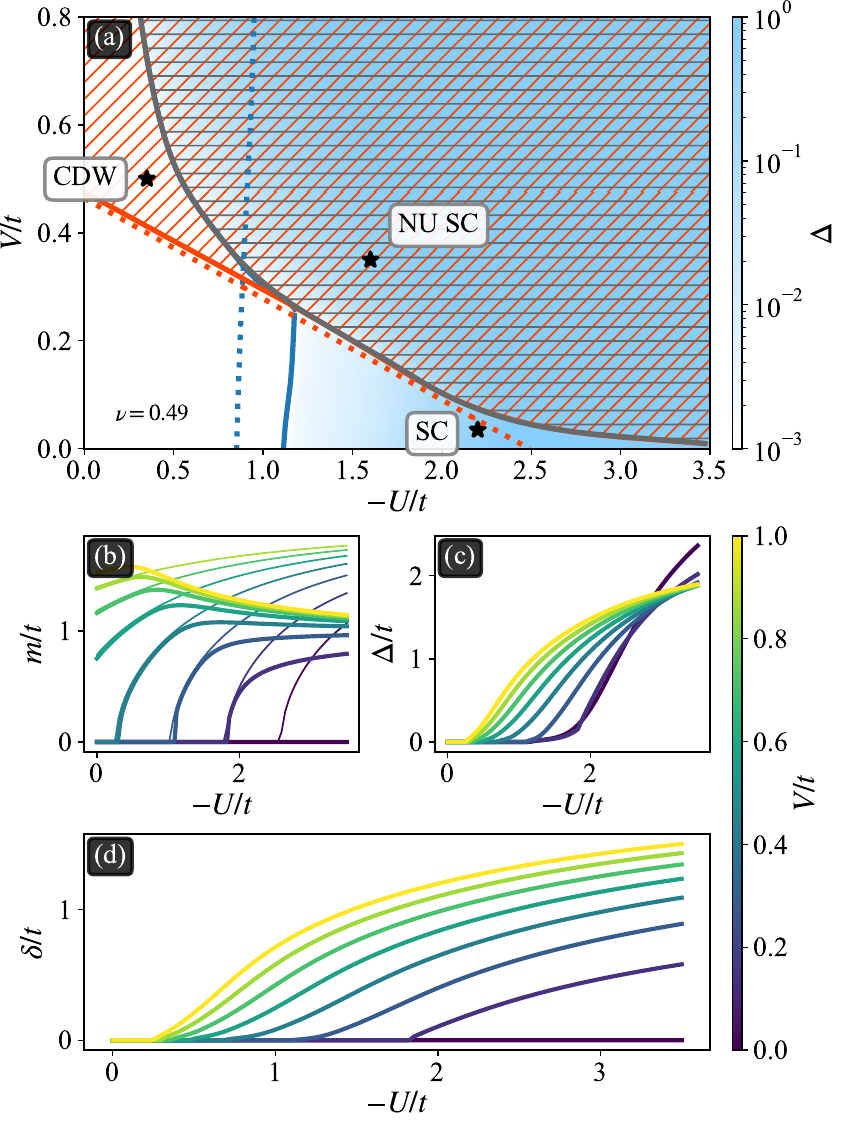}
\caption{%
Mean-field analysis of the interplay between charge density wave and spin singlet s-wave superconductivity in the honeycomb lattice for onsite attraction ($U<0$) and nearest-neighbor repulsion ($V>0$).
(a) Phase diagram showing the SC order parameter $\Delta_0$ (blue), regions with CDW order (red, striped), and regions with NU superconductivity (gray, striped). The transition lines (solid lines) indicate where the respective order parameter exceeds $10^{-3}t$. Transition lines when CDW order is artificially suppressed (blue, dotted) and when SC is suppressed (red, dotted) are also shown.
(b-d) Order parameters for varying onsite attraction $U$ at discrete values of nearest-neighbor repulsion (indicated by line colors).
Panel (b) shows the CDW order parameter $m$, (c) shows the order parameter $\delta$ for the pairing imbalance between the sublattices, while (d) shows the average $\Delta$ of the s-wave parameters in sublattices $A$ and $B$. The thin lines in panel (b) show the order parameter when SC is artificially suppressed.
The numerical calculations use $10^4$ points in reciprocal space, at filling $\nu=0.49$ and temperature $k_B T=10^{-3}t$.
}%
\label{fig3}
\end{figure}

For our microscopic model, we now solve the self-consistency relation for the mean fields numerically \cite{supmat}. 
We assume that the ground state preserves the original translational symmetries but can break other space group symmetries. 
We consider the mean-field electronic bands of $\overline{H}_0+\overline{H}_{P}$ [cf. \cref{fig2}], as well as the order parameters [cf.~\cref{fig3}].

The nearest-neighbor repulsion ($V>0$) alone favors the formation of a CDW \cite{PhysRevLett.97.146401,PhysRevLett.100.146404,PhysRevB.92.155137,RevModPhys.84.1067}. 
It reduces the symmetry group (${D_{6h}\to D_{3h}}$) by breaking the sublattice symmetry, and is characterized by the same order parameter $m=n_A-n_B$ we introduced earlier.
As a consequence, the Dirac cones in the electronic spectrum acquire a mass term, leading to a band gap as shown in \cref{fig2}(d).  
If we instead consider the case without nearest-neighbor repulsion, the attractive on-site interaction alone ($U<0$) allows pairs with s-wave orbital symmetry and singlet spin configuration to form. 
This is described by the pairing order parameter
$\Delta \equiv \sum_{\alpha} | \Delta_{\alpha}^0 |$
and leads to a hybridization between the electron and hole branches in the spectrum, see \cref{fig2}(e). 
Notably, the effect is small even for sizable interaction strengths, which is a result of the vanishing density of states near the Dirac point in the electronic spectrum. 

When both nearest-neighbor repulsion ($V>0$) and on-site attraction ($U<0$) are sizable, we find a third phase with both a CDW and s-wave singlet SC with the mean-field electronic band structure shown in \cref{fig2}(f). 
This phase is characterized by the order parameter
$\delta \equiv \left| |\Delta_{A}^0|- |\Delta_{B}^0| \right|$, 
that describes a sublattice imbalance in the s-wave pairing amplitude.
We see that the enhanced density of states near the band edges favors the formation of pairs. 
In what follows, we will investigate the phase diagram and the mechanism with which the CDW induces sublattice symmetry-broken pairing correlations.

In \cref{fig3}(a), we see the mean-field phase diagram showing the three symmetry-breaking phases we just introduced: the CDW when $V>0$ dominates (red), the unitary s-wave singlet superconductivity when $U<0$ dominates (blue), and a mixed phase with NU SC due to broken sublattice symmetry both in the charge and the pairing sectors (gray) when both interactions are sizable.
We find that the coexistence phase occupies the majority of phase space and always appears together with finite order parameters for the other phases. 
Furthermore, larger onsite interaction $U<0$ lowers the critical nearest-neighbor interaction $V_c$ at which a CDW occurs (red transition line). 
Increasing the nearest-neighbor interaction $V<V_c$ (below the onset of a CDW) disfavors the superconducting phase (blue transition line).
In \cref{fig3}(b--d), we find that (within our numerical resolution) the onset of the CDW and NU orders emerge as second-order phase transitions.

To investigate the nature of the coexistince phase, we repeat the self-consistent mean-field iteration without allowing for the CDW. 
We achieve this by enforcing $n_A=n_B$ in each step. 
In \cref{fig3}(a), we see that this constraint leads to a significantly modified phase diagram: while unitary s-wave pairing still occurs (blue dashed transition line), the sublattice-asymmetric NU pairing is no longer does, i.e., $\delta=0$. This means that that the NU SC relies on the presence of the charge density wave.
We also see that suppressing the CDW results in a shift in the (blue) transition line of the unitary SC order parameter $\Delta$ shown in \cref{fig3}(a). For small $V$, the constraint shifts the transition towards a smaller critical threshold $U_c$, which we attribute to a renormalized Fermi velocity (e.g., reduced density of states). At larger $V$, the threshold $U_c$ shifts towards larger values, which is in accordance with the observations made in the effective GL free energy densities, where we studied the coupling between the order parameters $m$ and $\delta$ [cf.~ \cref{eq:GL_coupling}]:
The CDW significantly enhances SC, allowing sizable pairings at much smaller on-site interactions $U<0$ (e.g. at $U_C\sim -0.6t$ for $V\sim0.6t$).

To summarize, we have shown that the existence of competing interactions in Dirac systems naturally leads to NU multiorbital SC states. 
In particular, we have shown that the interplay between SC and charge order fluctuations leads to a cooperative effect in which the appearence of NU SC order dramatically enhances the SC gap. 
Our results put forward competing interactions as a compelling mechanism giving rise to NU SC, and establish Dirac materials as paradigmatic system to realize unconventional multiorbital SC.

\begin{acknowledgments}
\textit{Acknowledgments}
We would like to thank Yuhao Zhao for helpful discussions. 
T.~M.~R.~W.\ acknowledges funding from the Swiss National Science Foundation (SNSF) through NCCR QSIT. 
M.~F.~H.\ and M.~S.\ are grateful for the financial support from the SNSF through Division II (No.~163186 and 184739).
J.~L.~L.\ acknowledges financial support from the Academy of Finland Projects No.~331342 and No.~336243,
and the Jane and Aatos Erkko Foundation.
\end{acknowledgments}

\bibliography{biblio.bib}


\onecolumngrid
\clearpage
\begin{center}
\textbf{\large Supplemental Material for \\ ``Non-unitary multiorbital superconductivity from competing interactions in Dirac materials''}
\end{center}
\setcounter{equation}{0}
\setcounter{figure}{0}
\setcounter{table}{0}
\setcounter{page}{1}
\makeatletter
\renewcommand{\thepage}{SM-\arabic{page}}
\renewcommand{\theequation}{S\arabic{equation}}
\renewcommand{\thefigure}{S\arabic{figure}}

\section{Ginzburg-Landau theory}\label{app:GL}

To derive the full Ginzburg-Landau (GL) free energy functional, we follow the general procedure outlined by Sigrist and Ueda \cite{RevModPhys.63.239}.
The superconducting (SC) state is described by the two complex-valued order parameters $\Delta_A$, $\Delta_B$, transforming according to the same irreducible representation (irrep) of the underlying point group $G$ of the crystal lattice.
We assume this irrep to be one-dimensional for our discussions.
The free energy functional needs to be constructed in such a way that it is invariant under point group operations (in particular including sublattice interchanging $A\leftrightarrow B$), $U(1)$ and time-reversal symmetry.
Therefore, all relevant second order terms are given by $\abs{\Delta_A}^2 +\abs{\Delta_B}^2$ and $(\Delta_A^*\Delta_B +c.c.)$.
In the same manner, we can find the fourth order terms to be $\abs{\Delta_A}^4 +\abs{\Delta_B}^4$, $\abs{\Delta_A}^2\abs{\Delta_B}^2$, $(\abs{\Delta_A}^2 +\abs{\Delta_B}^2)(\Delta_A^*\Delta_B +c.c.)$, $(\Delta_A^{*2}\Delta_B^2 +c.c.)$.
The total GL free energy density can be stated as 
\begin{align}
    f_{\text{SC}} &= \tilde{a}_1\left(\abs{\Delta_A}^2 +\abs{\Delta_B}^2\right) +\tilde{a}_2\left(\Delta_A^*\Delta_B^{\phantom{*}} +c.c.\right) \nonumber\\
    &\phantom{=} +\tilde{b}_1\left(\abs{\Delta_A}^4 +\abs{\Delta_B}^4\right) +\tilde{b}_2\abs{\Delta_A}^2\abs{\Delta_B}^2 +\tilde{b}_3\left(\abs{\Delta_A}^2 +\abs{\Delta_B}^2\right)\left(\Delta_A^*\Delta_B^{\phantom{*}} +c.c.\right) +\tilde{b}_4\left(\Delta_A^{*2}\Delta_B^2 +c.c.\right) \\
    &= a_1\left(\abs{\Delta_A}^2 +\abs{\Delta_B}^2\right) +a_2\abs{\Delta_A -\Delta_B}^2 \nonumber\\
    &\phantom{=} +b_1\left(\abs{\Delta_A}^2 +\abs{\Delta_B}^2\right)^2 +b_2\left(\abs{\Delta_A}^2 -\abs{\Delta_B}^2\right)^2 +b_3\left(\abs{\Delta_A}^2 +\abs{\Delta_B}^2\right)\abs{\Delta_A -\Delta_B}^2 +b_4\abs{\Delta_A -\Delta_B}^4,
\end{align}
where $a_1 = \tilde{a}_1 +\tilde{a}_2$, $a_2 = -\tilde{a}_2$, $b_1 = \tilde{b}_1/2 +\tilde{b}_2/4 +\tilde{b}_3 +\tilde{b}_4/2$, $b_2 = \tilde{b}_1/2 -\tilde{b}_2/4 +\tilde{b}_4/2$, $b_3 = -\tilde{b}_3 -2\tilde{b}_4$ and $b_4 = \tilde{b}_4$.
Including the charge-density wave (CDW) $f_{\text{CDW}} = \alpha m^2 +\beta m^4$ where $m$ is real and transforms according to the one-dimensional irrep of $G$ which flips sign under sublattice exchange, its coupling to the SC order parameter 
\begin{equation*}
    f_{\text{CPL}} = \lambda m(\abs{\Delta_A}^2 -\abs{\Delta_B}^2) +\zeta_1m^2\left(\abs{\Delta_A}^2 +\abs{\Delta_B}^2\right) +\zeta_2m^2\abs{\Delta_A -\Delta_B}^2
\end{equation*}
and introducing the parametrization $\Delta_{A, B} = \Delta \pm\delta/2$ yields
\begin{align}
    f &= 2a_1\abs{\Delta}^2 +\left(\frac{a_1}{2} +a_2\right)\abs{\delta}^2 +2\left(b_1 +b_2 +b_3\right)\abs{\Delta}^2\abs{\delta}^2 +b_2\left(\Delta^{*2}\delta^2 +c.c.\right) +4b_1\abs{\Delta}^4 +\left(\frac{b_1}{4} +\frac{b_3}{2} +b_4\right)\abs{\delta}^4 \nonumber \\
    &\phantom{=} +\alpha m^2 +\beta m^4 +\lambda m\left(\Delta^*\delta +c.c.\right) +2\zeta_1 m^2\abs{\Delta}^2 +\left(\frac{\zeta_1}{2} +\zeta_2\right)m^2\abs{\delta}^2.
\end{align}
The parameters $a_1 = a_1'(T -T_c^{\text{SC}})$, $a_1' > 0$, and $\alpha = \alpha'(T -T_c^{\text{CDW}})$, $\alpha' > 0$, are temperature dependent and $T_c^{\text{SC}}$ and $T_c^{\text{CDW}}$ denote the independent (=uncoupled) phase transition temperatures for the SC and the CDW, respectively.
Now, let us assume that $T_c^{\text{SC}} > T_c^{\text{CDW}}$ and $a_2 > 0$.
For $T > T_c^{\text{SC}}$, we find $\Delta = \delta = m = 0$.
Lowering the temperature slightly below the SC critical temperature, $T \lesssim T_c^{\text{SC}}$, results in $\Delta \neq 0$ and $\delta = m = 0$.
The magnitude of the SC order parameter $\Delta$ is determined by the corresponding GL equation
\begin{equation}
    0\overset{!}{=} 2a_1\Delta +8b_1\abs{\Delta}^2\Delta \iff \abs{\Delta}^2 = -\frac{a_1}{4b_1}.
\end{equation}
Lowering the temperature even further results in a second phase transition, with $\delta, m \neq 0$, whose transition temperature is determined by the corresponding linearized GL equations
\begin{align}
    0 &= \partial_{\delta^*}f \approx \left(\frac{a_1}{2} +a_2\right)\delta +2\left(b_1 +b_2 +b_3\right)\abs{\Delta}^2\delta +2b_2\Delta^2\delta^* +\lambda m\Delta \nonumber \\
    0 &= \partial_{m}f \approx 2\alpha m +\lambda\left(\Delta^*\delta +c.c.\right) +4\zeta_1\abs{\Delta}^2m.
\end{align}
The coupling term $f_{\text{CPL}}$ suggests that $\Delta$ and $\delta$ do have a relative phase of $0$ or $\pi$ depending on the sign of $\lambda$ and therefore both can be rendered to be real such that the linearized GL equations for $\delta$ and $m$ can be rewritten as
\begin{equation}
    0 = \begin{pmatrix} \frac{a_1}{2} +a_2 +2\left(b_1 +2b_2 +b_3\right)\Delta^2 & \lambda\Delta \\ \lambda\Delta & \alpha +2\zeta_1\Delta^2 \end{pmatrix}\begin{pmatrix} \delta \\ m \end{pmatrix}.
\end{equation}
The phase transition temperature is determined by the highest temperature such that the determinant of the $2\times 2$-matrix is vanishing
\begin{equation}
    0 = \left[\left(\frac{1}{2} -\frac{b_1 +2b_2 +b_3}{2b_1}\right)a_1(T) +a_2\right]\left(\alpha(T) -\frac{\zeta_1a_1(T)}{2b_1}\right) +\frac{\lambda^2 a_1(T)}{4b_1}.
\end{equation}
The symmetric off-diagonal couplings, $\lambda\Delta$, increase the transition temperature above the bare ones for $\delta$ and $m$. 
In particular we observe $T_c > T_c^{\text{CDW}}$.

For completeness, we note that an analogous discussion can be applied to the case $T_c^{\text{CDW}} > T_c^{\text{SC}}$.

\section{Intuitive weak-coupling argument for cooperation between SC and CDW}

\begin{figure}[t]
\centering
\includegraphics[width=\columnwidth]{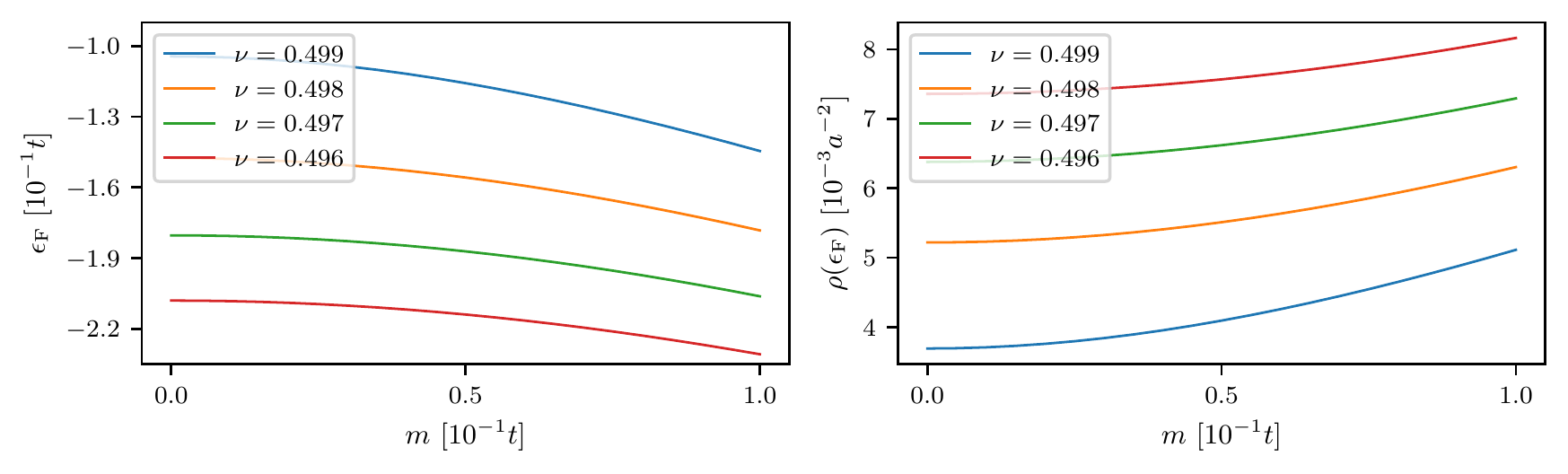}
\caption{%
Fermi level $\epsilon_{\text{F}}$ (a) and DOS at the Fermi level $\rho(\epsilon_{\text{F}})$ (b) as a function of charge imbalance $m$ for different filling factors $\nu$ close to charge neutrality $\nu \lesssim 1/2$.
}
\label{fig:DOS}
\end{figure}

Within the weak-coupling theory of SC, the SC gap $\Delta$ can be related to the attractive interaction strength $U$ as well as to the density of states (DOS) at the Fermi level $\rho(\epsilon_{\text{F}})$ via
\begin{equation}
    \Delta \sim e^{-\frac{1}{\abs{U}\rho(\epsilon_{\text{F}})}}.
\end{equation}
Here, we aim to provide an intuitive argument why the CDW order benefits the SC order.
For this, we assume a situation where we are close to half-filling $\nu \lesssim 1/2$, i.e.~a chemical potential slightly below the Dirac point, and show that the CDW order introduces a finite mass $m$, increasing the DOS $\rho(\epsilon_{\text{F}})$ and therefore also the SC gap $\Delta$.

At first, we determine the Fermi level $\epsilon_{\text{F}}$ as a function of the filling factor $\nu < 1/2$ for the single-particle tight-binding band structure of graphene
\begin{equation}
    \epsilon_{\alpha, s}(\K) = (\pm^{\alpha}) \sqrt{t^2\left[3 + 2\left(\cos(\K\cdot\bs{a}_1) +\cos(\K\cdot\bs{a}_2) +\cos(\K\cdot(\bs{a}_1 -\bs{a}_2))\right)\right] +m^2},
\end{equation}
where $\alpha$ is the sublattice index and $s$ the spin index.
At $T = 0$ the Fermi level is determined by the equation
\begin{equation}
    2N\nu = \sum_{\K}\Theta(\epsilon_{\text{F}} -\epsilon_{\alpha=-, s}(\K))
\end{equation}
which needs to be solved numerically for $\epsilon_{\text{F}}(\nu, m)$, see \cref{fig:DOS}(a).
$\Theta$ denotes the Heaviside function and $N$ is the total number of lattice sites.

Next, for $\nu \lesssim 1/2$, we approximate the band structure by a linear Dirac band structure $\tilde{\epsilon}(\K)$ and compute the Fermi wave vector $k_{\text{F}}$ (relative to either of the two Dirac points) corresponding to the Fermi level $\epsilon_{\text{F}}$ via
\begin{equation}
    \epsilon_{\text{F}} \overset{!}{=} \tilde\epsilon(k_{\text{F}}) = -\sqrt{(3/2at)^2k_{\text{F}}^2 +m^2} \iff k_{\text{F}}(\nu, m) = \sqrt{\frac{\epsilon_{\text{F}}(\nu, m)^2 -m^2}{(3/2at)^2}}.
\end{equation}
From this we obtain the Fermi velocity as
\begin{equation}
    v_{\text{F}} = \frac{1}{\hbar}\partial_{\abs{\K}}\tilde\epsilon(\K)\rvert_{k_{\text{F}}} = -\frac{1}{\hbar}\frac{2(3/2at)^2k_{\text{F}}}{\sqrt{(3/2at)^2k_{\text{F}}^2 +m^2}}
\end{equation}
and can finally compute the DOS at the Fermi level
\begin{equation}
    \rho(\epsilon_{\text{F}}) = \frac{1}{4N}\sum_{n, \K}\delta_{\epsilon_{\text{F}}, \epsilon_n(\K)} = \frac{a^2}{16\pi^2}\sum_n\int_{0}^{2\pi/a}dk_1\int_{0}^{2\pi/a}dk_2\delta(\epsilon_{\text{F}} -\epsilon_n(\K)),
\end{equation}
where the additional integral normalization $I_N$ which is introduced by replacing the sums by integrals is determined by the expression $\sum_{\K\in BZ}1 =1/{I_N}\int_0^{2\pi/a}dk_1\int_0^{2\pi/a}dk_2 1$.
Using the fact that the bands are spin degenerate, that we have two Dirac cones in the Brillouin zone, and that $\epsilon_{\text{F}} < 0$ ($\nu < 1/2$), we can rewrite the DOS as
\begin{equation}
    \rho(\epsilon_{\text{F}}) = \frac{a^2}{4\pi^2}\int_{A_{\text{Dirac}}}d^2\K\delta(\epsilon_{\text{F}} -\tilde\epsilon(\K)),
\end{equation}
where $A_{\text{Dirac}}$ is an area completely containing the Fermi surface.
Switching to polar coordinates centered around the chosen Dirac cone and using a Dirac-$\delta$ identity (under the assumption that $\tilde\epsilon$ is only depending on the radius) yields
\begin{equation}
    \rho(\epsilon_{\text{F}}) = \frac{a^2}{2\pi}\int_{k_{\text{F}}-\gamma}^{k_{\text{F}}+\gamma}dk \frac{k\delta(k -k_{\text{F}})}{\hbar \abs{v_{\text{F}}}} = \frac{a^2}{2\pi}\frac{k_{\text{F}}}{\hbar \abs{v_{\text{F}}}}
\end{equation}
for a small $\gamma \gtrsim 0$.
\cref{fig:DOS}(b) shows that the DOS at the Fermi level is monotonously increasing in $m$ and thus SC is enhanced by the additional charge order.

\section{Mean-field analysis of interactions in the tight-binding model}

\subsection{General strategy}

For the reader's convenience, we briefly summarize the mean-field approximation used in this work. We note that our tight-binding Hamiltonian is of the generic form
\begin{align}
H = H_0 + V = \sum_{\nu} \epsilon_{\nu} c_{\nu}^{\dagger} c_{\nu} +
   \sum_{\mu \neq \nu} t_{\mu \nu} c_{\mu}^{\dagger} c_{\nu} + \frac{1}{2} 
   \sum_{\mu \neq \nu} V_{\mu \nu} c^{\dagger}_{\nu} c^{\dagger}_{\mu} c_{\mu}
   c_{\nu},
\end{align}
where $\cd_\nu$ ($\cpd_\nu$) creates (destroys) single-particle state $\ket{\nu}$ with quantum numbers $\nu$, we have on-site energies $\epsilon_\nu$, transfer amplitudes $t_{\mu\nu}$, and interaction strengths $V_{\mu\nu}$. Assuming that quantum-statistical fluctuations around expectation values $G_{\mu\nu}=\langle \cd_\mu \cpd_\nu \rangle$ and $F_{\mu\nu}=\langle \cd_\mu \cd_\nu \rangle$ are small (which holds away from phase transitions), we can approximate interaction terms using the substitution
\begin{align} \label{eq:meanfield_approximation}
    c^{\dagger}_{\nu} c^{\dagger}_{\mu} c_{\mu} c_{\nu} & \approx G_{\nu \nu}
  c^{\dagger}_{\mu} c_{\mu} + G_{\mu \mu} c^{\dagger}_{\nu} c_{\nu} - G_{\nu
  \mu} c^{\dagger}_{\mu} c_{\nu} - G_{\mu \nu} c^{\dagger}_{\nu} c_{\mu} -
  G_{\nu \nu} G_{\mu \mu} + | G_{\mu \nu} |^2 
  + F_{\mu \nu} c_{\mu} c_{\nu} + F_{\mu \nu}^{\ast}
  c^{\dagger}_{\nu} c^{\dagger}_{\mu} - | F_{\nu \mu} |^2,
\end{align}
which leads to both Hartree- and Fock-type corrections due to interactions, as well possible pairing correlations.

The statistical expectation value is defined through $\langle \cdots \rangle = Z^{-1} \tr(e^{-\beta (H-\mu N)} \cdots)$ with partition function $Z=\tr(e^{-\beta (H-\mu N)})$, where $N$ is the particle number operator, and $\mu$ is the chemical potential. If $H$ is quadratic (i.e., non-interacting with or without pairing terms), we can find the eigenbasis \cite{xiao2009theory} and evaluate the expectation value. 
\emph{However}, if $H$ contains interactions, we can use the mean-field substitution \cref{eq:meanfield_approximation} to approximate it with a non-interacting (quadratic) Hamiltonian $H^{\text{MF}}(G,F)$. In this case, the mean-fields $G$ and $F$ have to be calculated self-consistently through the defining relations
\begin{align}
    G_{\mu\nu} = \langle \cd_\mu \cpd_\nu \rangle = \frac{1}{Z_{\text{MF}}} \tr(e^{-\beta (H^{\text{MF}}-\mu N)} \cd_\mu \cpd_\nu), &&
    F_{\mu\nu} = \langle \cd_\mu \cd_\nu \rangle = \frac{1}{Z_{\text{MF}}} \tr(e^{-\beta (H^{\text{MF}}-\mu N)} \cd_\mu \cd_\nu),
\end{align}
where $G_{\mu\nu}$ ($F_{\mu\nu}$) appear both on the left-hand \emph{and} on the right-hand side through $H^{\text{MF}}(G,F)$. This problem is usually solved numerically with the objective to find the solution that produces the lowest overall groundstate energy. Typically, one additionally enforces the constraint on $\mu$ that the filling $\nu$ (and hence the particle number) should remain constant.

\subsection{Mean-field terms in our model}

For our numerical tight-binding study, we have implemented a package that automatically calculates and includes all possible mean-field terms for any given interaction potential without the need to define each of them explicitly (which is prone to human error). However, since it can be insightful to examine the terms explicitly, we state them here for the interested reader.

We apply the mean-field approximation to the Hamiltonian in \cref{eq:extended_hubbard} and assume translational symmetry on the ground state to find the ``normal'' mean-field terms
\begin{align}
  H_{\textrm{int}}^{\textrm{MF},N} &= \sum_{i,\alpha,s} (-\dmu_{\alpha s}) \, n_{\ia s} + \lambda_{\alpha s} \, \cd_{\ia s} \cpd_{\ia \overline{s}} + 
  \sum_{\langle iA, jB\rangle,s} \dt_{ss'} \, c_{iA s}^{\dagger} c_{jB s'}^{\pdagger} + \textrm{h.c.},
\end{align}
where we introduced the mean-field contributions to the local chemical potential, local spin flips, and non-local hoppings, i.e.,
\begin{align}
  \dmu_{\alpha s} = -U \langle n_{\ia \overline{s}} \rangle + V \langle n_{i \overline{\alpha}} \rangle, &&
  \lambda_{\alpha s} &= - U \langle c^\dagger_{\ia s} c^\pdagger_{\ia \overline{s}} \rangle^*, &&
  \dt_{ss'} = - V \langle \cd_{i A s} \cpd_{j B s'} \rangle^*. 
\end{align}
The mean field expansion also introduces ``anomalous'' particle--particle pairing terms, which can be decomposed into singlet ($\nu=0$) and triplet ($\nu=x,y,z$) components, i.e.,
\begin{align}
  H_{\textrm{int}}^{\textrm{MF},A}  = \frac{U}{2} \sum_{i,\alpha} F_{i\alpha,i\alpha}^{0\;\ast} \, c^\dagger_{\ia s} (-i \sigma^y)_{ss'}^{\pdagger} c^\dagger_{\ia s'} 
  + \frac{V}{2} \sum_{\langle iA,jB\rangle,\nu} F_{iA,jB}^{\nu\; \ast} \, c^\dagger_{i A s} (-i \sigma^y\sigma_\nu)_{ss'} c^\dagger_{j B s'}  
  + \textrm{h.c.},
\end{align}
where $F_{i\alpha,j\beta}^{\nu} \equiv (-i \sigma_y\sigma_\nu)_{ss'} F_{i\alpha,j\beta}$ are singlet and triplet components of the pairing mean field.
Rotational symmetry in the spin sector allows to fix $\Delta_{i\alpha,j\beta}^{x,y}=0$.
Note that in the main text, we (silently) neglected the triplet contribution. However, it \emph{is} included in our numerical study and we found it to be insignificant for the parameter regime studied in this work.

\clearpage

\end{document}